\documentclass[preprint,superscriptaddress]{revtex4}

\usepackage{graphicx}
\usepackage{bm}
\usepackage{amsmath}
\usepackage{amssymb}
\usepackage{hyperref}
\usepackage{natbib}
\usepackage{amsfonts}

\begin{document}

\newcommand{\inthalf}{\int\limits_0^\infty}
\newcommand{\intpi}{\int\limits_0^{\pi/2}}
\newcommand{\pdd}[1]{\frac{\partial}{\partial #1}}

\title{Electrostatics of Graphene: Charge Distribution and Capacitance}

\author{Zhao Wang}
\email{wzzhao@yahoo.fr}
\affiliation{EMPA - Swiss Federal Laboratories for Materials Testing and Research, Feuerwerkerstrasse 39, CH-3602 Thun, Switzerland.}

\author{Robert W. Scharstein}
\affiliation{Electrical Engineering Department, University of Alabama, 317 Houser Hall, Tuscaloosa, AL 35487-0286, USA.}

\begin{abstract}
The distribution of net electric charge in graphene is investigated, using both a constitutive atomic charge-dipole interaction model and an approximate analytical solution to Laplace's equation. We demonstrate a strong size dependence of the charge distributions in finite-size, infinitely-long and multi-layered rectangular graphene sheets, respectively. We found that the charge density can be naturally enhanced up to 13 times at graphene's geometry edges. This edge charge enhancement effect becomes more significant when the length or the width of graphene increases. The charge enhancement ratio is found to follow a linear relationship with the number of layers. These results can be used to understand the newly experimentally observed electron emission, charge impurity and chemical doping phenomena in 2-dimension nanostructure.
\end{abstract}


\maketitle

\section{Introduction}

Graphene is an hexagonally coordinated semi-metallic or zero-gap-semiconducting carbon layer that crystallizes in $sp^2$ structure. Its unusual band structure \cite{CastroNeto2009,Beenakker2008} leads to unique electronic properties such as high charge carrier mobility \cite{Hwang2007d,Morozov2008a,Fogler2008} and tunable electron band gap by external electromagnetic fields \cite{Son2006b,Yang2007c,Han2007a,Castro2007,Novikov2007a,Yan2007a} or by chemical doping.\cite{Duplock2004,Biel2009} These make graphene an ideal candidate for a number of applications in nanoelectronics.\cite{Westervelt2008} Recently, it has been shown by electrostatic force microscopy experiments that electric charges can be stored in carbon nanotubes for a period of time.\cite{Paillet2005,Zdrojek2005,Jespersen2005} It would not be surprising if graphene exhibits similar properties, because of the large similarity in its atomic structures to that of nanotubes.\cite{Charlier2007} The investigation on electronic properties of graphene doped with net charges is of importance for understanding the charge impurities \cite{Morozov2006,Yan2008,Rossi2008} or chemical doping \cite{Duplock2004,Biel2009} in graphene, and may as well lead to further applications in charge storages,\cite{Cui2002} sensors \cite{Robinson2008} and field-emission devices \cite{Bonard2002,Purcell2002} based on graphene. 

The question about the exact solution to the charge distribution in an arbitrary-shaped conducting surface has been raised since the theoretical basis of electromagnetism was established by J. C. Maxwell in 1873.\cite{Maxwell1873} For graphene, the electrostatic problem of finding the charge distribution on the surface of a conducting, rectangular plate requires the solution to Laplace's equation in the three-dimensional geometry. Unfortunately, even such a seemingly simple problem has no exact solution: The number of exact solutions in three-dimensions is still limited to specialized geometries that exhibit a great deal of symmetry\cite{Jacksonbook1975,Morse1953,Bottcher1952}. In this paper, the charge distribution was first calculated using a numerical atomic charge-dipole model. Comparisons are then made between the result of a simple, one-term variational approximation and a numerical solution, which is found to give a more accurate description of the edge atoms.

The remainder of this paper is organized as follows. In Sec. (\ref{Sec2}) we describe first our numerical method and then we demonstrate a variational approach to solve Laplace's equation and calculate the charge distribution on a conducting rectangular plate. The results of both numerical and analytical models are compared and discussed in Sec. (\ref{Sec3}). We draw a conclusion in Sec. (\ref{Sec4}). Mathematical details of the one-term variational approximation 
are summarized in the Appendix.

\newpage

\section{Theory}
\label{Sec2}

\subsection{Numerical method}
\label{Sec2a}
In this work, the charge density was first calculated using a computational atomic charge-dipole model,\cite{mayer-07-01} in which each atom is modeled as a polarizable sphere with a net charge $q$ and an induced dipole $\bm{p}$. The equilibrium distribution of the charges and the induced dipoles in graphene is determined by minimizing the total electrostatic potential energy of the system

\begin{equation}
\label{eq:1}
\begin{array}{cl}
U^{elec}= & \sum\limits_{i=1}^N{q_i(\chi_i+V_i)}-\sum\limits_{i=1}^N{\bm{p}_i\cdot\bm{E_i}}+
\frac{1}{2}\sum\limits_{i=1}^N{\sum\limits_{\substack{j=1}}^N{q_i T^{i,j}_{q-q} q_j}}\\
 & -\sum\limits_{i=1}^N{\sum\limits_{\substack{j=1}}^N{\bm{p}_i \cdot \bm{T}^{i,j}_{p-q} q_j}}
-\frac{1}{2}\sum\limits_{i=1}^N{\sum\limits_{\substack{j=1}}^N{\bm{p}_i \cdot \bm{T}^{i,j}_{p-p} \cdot \bm{p}_j}}
\end{array}
\end{equation}

\noindent where $\chi$ is the electron affinity, and where $V$ and $\bm{E}$ are the external potential and electric field, respectively. The electrostatic interacting tensors are $T$ and $\bm{T}$, which are  regularized by a Gaussian distribution \cite{ZW2009}. Requiring the derivatives of $U^{elec}$ with respect to $q$ and $\bm{p}$ to be zero, Eq. (\ref{eq:1}) yields 

\begin{equation}
\label{eq:2}
\begin{array}{c}
\left\{ \begin{array}{ll}
\sum\limits_{j=1}^N{\bm{T}^{i,j}_{p-p}\otimes\bm{p}_{j}}+\sum\limits_{j=1}^N{\bm{T}^{i,j}_{p-q}q_{j}}=-\bm{E_i}\\
\sum\limits_{j=1}^N{\bm{T}^{i,j}_{p-q}\cdot\bm{p}_{j}}+\sum\limits_{j=1}^N{T^{i,j}_{q-q}q_{j}}+\lambda = -(\chi_i+V_i)\\
\sum\limits_{j=1}^N{q_{j}}=Q^{tot}
\end{array}\right.\\
\forall{i=1,...,N}
\end{array}
\end{equation}

\noindent where $\lambda$ is the Lagrange multiplier which is related to the chemical potential of the molecule.

This numerical model is an extension of the atomic dipole interaction theories \cite{Applequist-72,olson-78,Jensen2002}. It has recently been parameterized for $sp^{2}$ carbon systems \cite{mayer-05-01,Mayer2005c} and validated through electrostatic force microscopy charge injection experiments.\cite{zhaowang-08-01} The graphene structure is relaxed by means of an atomic energy-optimization method using the adaptive interatomic reactive empirical bond order (AIREBO) potential function,\cite{Stuart2000a} which has specially been parameterized for $sp^{2}$ hydrocarbon systems.\cite{Ni2002,Shenoy2008,ZW200903,zhaowang-07-03,zhaowang-07-02} 

\subsection{Analytic approximation}

A simple analytical approximation for the solution to Laplace's equation is now constructed. In three physical dimensions, consider an infinitesimally thin conducting surface of rectangular shape $L_{x}$-by-$L_{y}$ that is held to the potential $V_0$, with the reference or zero potential at infinity. The rectangle lies in the $z=0$ plane and is the region $|x|<L_{x}/2$, $|y|<L_{y}/2$. We seek a solution to Laplace's equation
\begin{equation}
\label{eq:ana1}
\left(\frac{\partial^2}{\partial x^2}+\frac{\partial^2}{\partial y^2}+
\frac{\partial^2}{\partial z^2}\right)\psi(x,y,z)=0
\end{equation}
subject to the boundary conditions
\begin{equation}
\label{eq:ana2}
\begin{array}{c}
\left\{ \begin{array}{lr}
\psi(x,y,0)=V_0 \qquad & (|x|<L_{x}/2,|y|<L_{y}/2) \\
\psi(x,y,z)\to 0 & \mbox{  as  } \sqrt{x^2+y^2+z^2}\to\infty
\end{array}\right.\\
\end{array}\,.
\end{equation}

Based on experience with the exact solution for the two-dimensional strip problem,\cite{Morse1953,Jacksonbook1975} we propose a one-term approximation for the (scaled) charge density  
\begin{equation}
\label{eq:ana7}
\pdd{z}\psi(x,y,0^+)=\left\{\begin{array}{ll} \displaystyle{\frac{-k}{\sqrt{1-(2x/L_{x})^2}\sqrt{1-(2y/L_{y})^2}}},
 & \mbox{$|x|<L_{x}/2$ and $|y|<L_{y}/2$} \\ 0, & \mbox{$|x|>L_{x}/2$ or $|y|>L_{y}/2$}
\end{array}\right.
\end{equation}
where the single unknown is the constant $k$. This functional form is simply a factorization
of the rectangular distribution into two edge-condition forms that apply to the classical
two-dimensional strip problem. Its suitability is not a priori known, but ultimate
comparison with independent moment method \cite{HarringtonMM} calculations shows that it is
quite respectable.
The one-term Galerkin procedure in the Appendix produces an approximation for the constant 
\begin{equation}
\label{eq:ana18bis}
k=\frac{2V_0}{L_{x}I(L_{x}/L_{y})}
\end{equation}
in terms of an integral function $I(L_x/L_y)$ of Eq. (\ref{eq:ana17}) which is evaluated numerically. The resultant surface charge density is then
\begin{equation}
\label{eq:ana19}
\frac{\sigma(x,y)}{\epsilon_0}=-2\pdd{z}\psi(x,y,0^+)= \frac{2k}{\sqrt{1-(2x/L_{x})^2}\sqrt{1-(2y/L_{y})^2}} \,\,.
\end{equation}

Integration of Eq. (\ref{eq:ana19}) gives the total charge
\begin{equation}
\label{eq:ana20}
\frac{Q}{\epsilon_0}=4\int\limits_0^{L_{x}/2} dx\int\limits_0^{L_{y}/2} dy \,\frac{\sigma(x,y)}{\epsilon_0}
=4\cdot 2k\left(\frac{\pi L_{x}}{4}\right)\left(\frac{\pi L_{y}}{4}\right)
=\frac{\pi^2 L_y V_0}{I(L_x/L_y)} 
\end{equation}
induced on the rectangular plate when held to the potential $V_0$.

\begin{figure}[ht]
\centerline{\includegraphics[width=13cm]{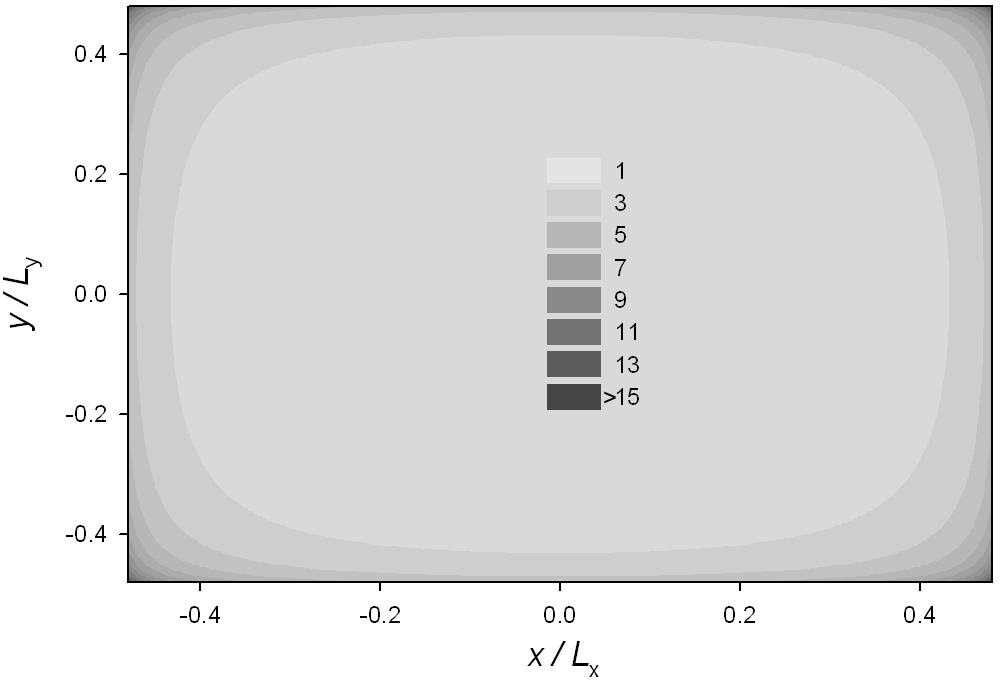}}
\caption{\label{fig:0}
Surface charge density ($\sigma$) normalized to that in the plate center. 
} 
\end{figure}

Here Fig. \ref{fig:0} shows the charge enhancement effect in a rectangular plate exhibited by Eq. (\ref{eq:ana19}). In addition to the surface charge distribution, it is instructive to also consider the capacitance $C=Q/V_0$, a global electrostatic characterization of the plate. 
In the present notation, the capacitance per unit width is
\begin{equation} 
\label{eq:ana21}
\frac{C}{L_{y}}=\frac{\pi^{2} \epsilon_{0}}{I(L_x/L_y)}  \,\,.
\end{equation}
In the limiting case of a square, $L_{x}=L_{y}$ and the numerics give $I(1)=2.176$ and the subsequent normalized capacitance is (in SI units) $C/L_{y}=40.159\,$(pF/m), about 1.6\% lower than the value $40.811\,$(pF/m) that has been published by Goto \textit{et al.}\cite{Gotoi1992}

\newpage

\section{Results and Discussions}
\label{Sec3}
\subsection{Finite-size graphene}
\begin{figure}[ht]
\centerline{\includegraphics[width=12cm]{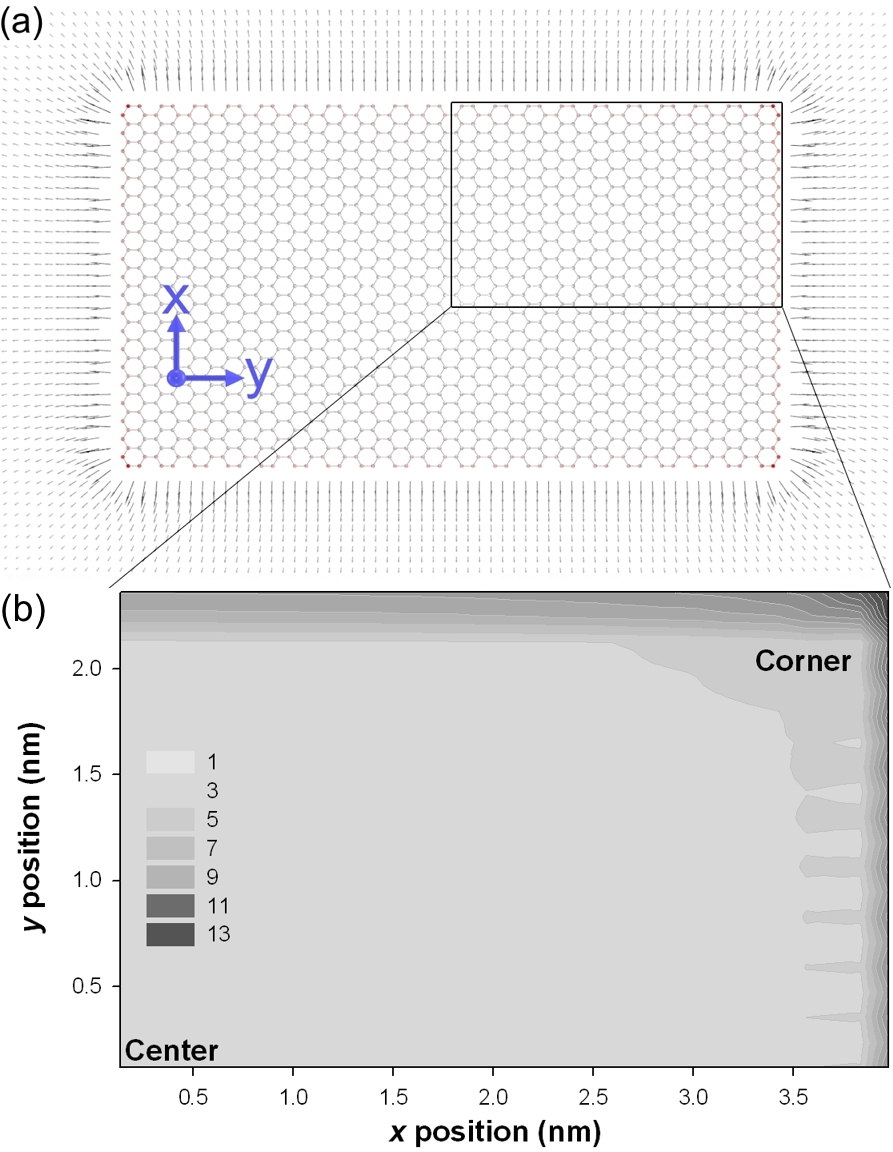}}
\caption{\label{fig:1}
(Color online) (a) Representative atomic diagram of net electric charges density in a graphene sheet ($L_{x} \times L_{y} = 5 \times 8$ nm). The color scale of atom is proportional to the charge density. The dark arrows stand for the electric fields induced by the net charges. The armchair edge is along the $x$ axis, and the zigzag one is in $y$ direction. (b) Atomic charge density normalized to that in the graphene center. 
} 
\end{figure}

We start with mono-layered graphene which contains a quantity of net electric charge. Fig. \ref{fig:1} (a) shows the charge distribution in a finite-size graphene sheet, calculated using the numerical method (Sec. (\ref{Sec2a})). As expected, typical charge and electric field enhancement effects \cite{zhaowang-08-01} were observed near the corners and the edges. To compare these results with those predicted by Eq. (\ref{eq:ana19}), we show the atomic charge density in a quarter of the graphene sheet in Fig. \ref{fig:1} (b). Comparison between the contours of charge density in Fig. \ref{fig:0} and Fig. \ref{fig:1} (b) shows a general agreement between the analytical model and the numerical calculation. The lack of smooth trend of the contours in Fig. \ref{fig:1} (b) is due to the different configurations of atoms at armchair and zigzag edges and the discontinuity of our discrete model. It is found that the Eq. (\ref{eq:ana19}) becomes no more accurate for the atoms near the edges, because the assumption of infinitely sharp edges in the analytical solution is not valid in atomic scale. After our simulation results, the charge density (at the corner) can be enhanced up to $14$ times over that in the center of graphene. We note that the charge density can also slightly vary if the graphene is substitutionally doped by net charges (e.g. some electrons per 100 atoms), since the graphene band structure can be significantly modified in highly charged graphene.\cite{Keblinski2002} Moreover, these results also imply that the influence of induced dipoles on the charge distribution is minor.

\begin{figure}[ht]
\centerline{\includegraphics[width=13cm]{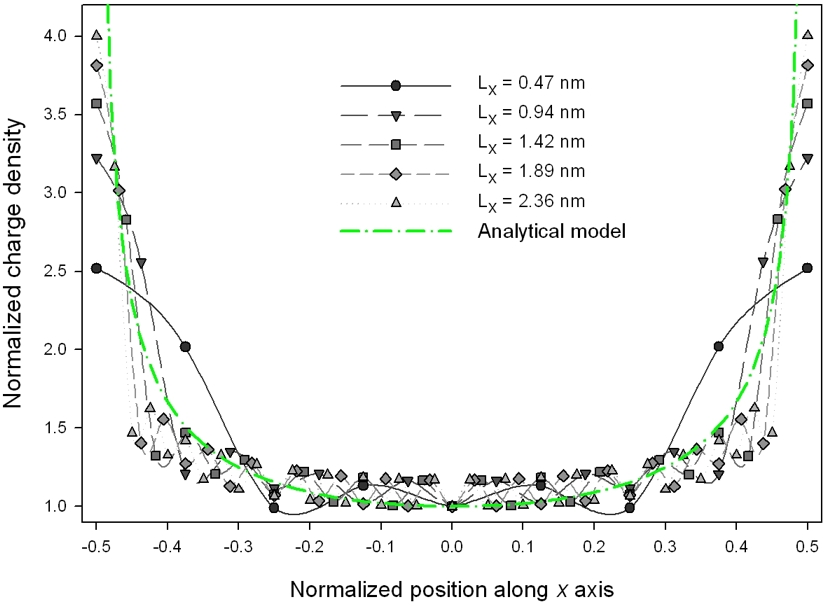}}
\caption{\label{fig:2}
Charge profile along the $x$ axis in 5 graphene sheets ($L_{y}=0.23$ nm) with different lengths $L_{x}$. The charge density is normalized to that in the center of graphene. The $x$ position is normalized to $L_{x}$. The symbols represent the simulation data and the dot-line curve stands for the analytical solution from Eq. (\ref{eq:ana19}).
} 
\end{figure}

We next consider the size effect on the charge distribution. Our results show a strong dependence of the charge enhancement on the graphene length $L_{x}$ (Fig. \ref{fig:2}). We can see that the fluctuation of the charge profile due to the edge state becomes less significant with the increasing length, and that the charge enhancement at the edges is stronger for the longer sheets. This behavior is similar to that in carbon nanotubes.\cite{Keblinski2002} 

\begin{figure}[ht]
\centerline{\includegraphics[width=13cm]{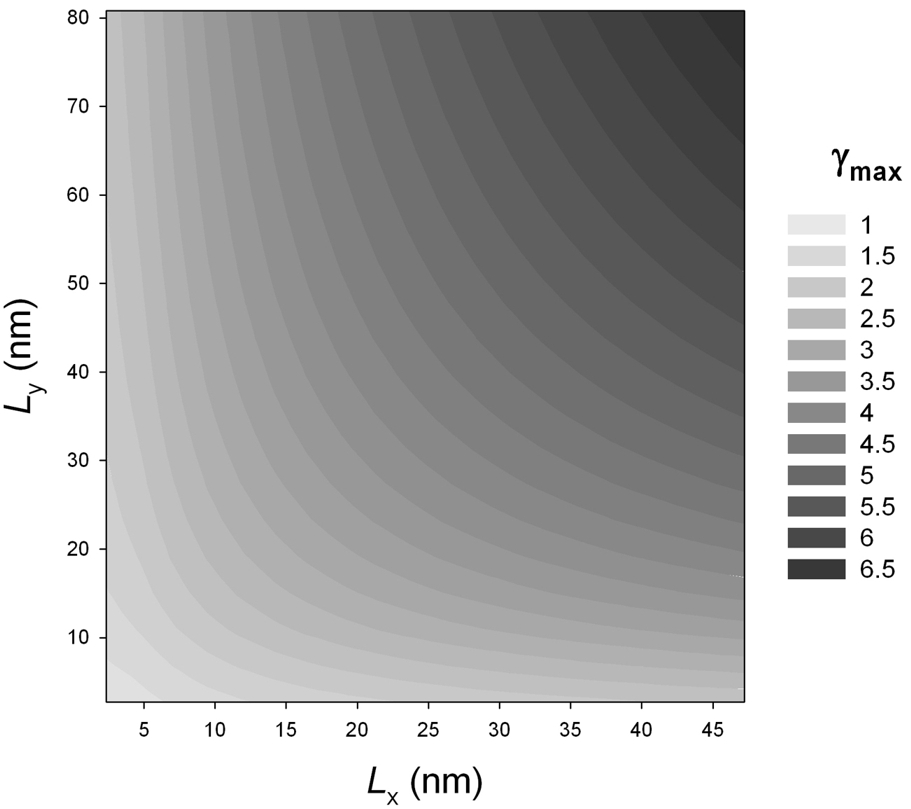}}
\caption{\label{fig:3}
Maximal charge enhancement factor $\gamma_{max}$ vs. the graphene size. $L_{x}$ and $L_{y}$ stand for the dimensions of graphene along the armchair and the zigzag edges, respectively. 
} 
\end{figure}

We now define the maximal charge enhancement ratio $\gamma_{max}$ as the maximal atomic charge density normalized to the average over the whole graphene. It is one of the key parameters for the potential applications of graphene in field emission devices \cite{Bonard1998}. To show a complete correlation between $\gamma_{max}$ and the dimensions of graphene, we plot numerical calculation data of $\gamma_{max}$ in Fig. \ref{fig:3} as a function of $L_{x}$ and $L_{y}$. We can see that $\gamma_{max}$ increases with either $L_{x}$ or $L_{y}$, and that $\gamma_{max}$ gets higher for $L_{x} \approx L_{y}$. It is found that the increase of charge enhance ratio is more rapid in small size graphene sheets. 

\subsection{Infinitely-long graphene}
\begin{figure}[ht]
\centerline{\includegraphics[width=13cm]{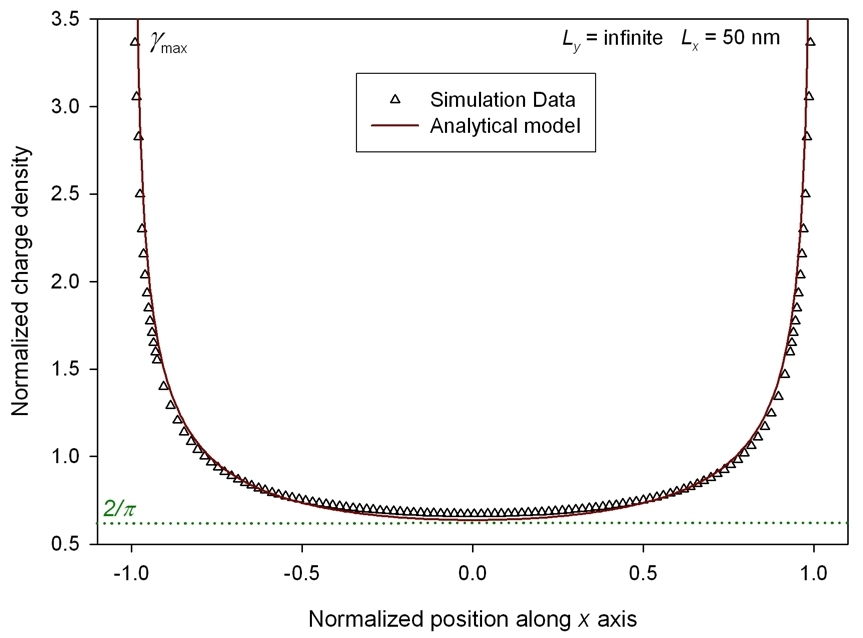}}
\caption{\label{fig:4}
Profile of charge density (normalized to the average) along the $x$ axis in a charged graphene sheet ($L_{x}=50$ nm). The periodic condition is applied along $y$ axis hence $L_{y} \rightarrow \infty$. The symbols represent the simulation data and the curve stands for the classical solution from Ref. \cite{Morse1953}.
} 
\end{figure}

To generalize our results for the graphene size usually used in experiments, it is particularly interesting to investigate the charge distribution in infinitely long graphene sheets or strips. The question of charge distribution becomes a 2D standard problem in elliptical coordinates.\cite{Morse1953} The exact solution to Laplace's equation for the strip conductor of width $L_x$
held to the potential $V_0$ gives the surface charge density
\begin{equation} 
\label{eq:ana22}
\sigma(x)=\frac{4\epsilon_{0}V_{0}}{L_{x} \ln{(4\rho_{0}/L_{x})}\sqrt{1-(2x/L_{x})^{2}}},
\end{equation}
where $\rho_0$ is the (large) radial distance where the potential vanishes. This is the 
\lq\lq ground\rq\rq\  or reference surface that cannot be at infinity because of the logarithmic singularity in the 2D potential problem. Integration of the surface charge density yields the capacitance per unit length
\begin{equation} 
\label{eq:ana23}
C_\ell=\frac{2 \pi \epsilon_{0}}{\ln(2 \rho_{0} / L_{x})}\,.
\end{equation}
Here we show the normalized charge profile in a graphene sheet with applied periodic condition in width ($y$) direction in Fig. \ref{fig:4}, which highlights a good agreement obtained between the numerical calculations and the analytical solution. Also, compared to Fig. \ref{fig:2}, it can be seen that the fluctuation on the charge density curves due to the open-edge effects \cite{Liu2009} completely disappeared in the infinite-long graphene.\cite{Yang2008d,Abanin2006} 

\begin{figure}[ht]
\centerline{\includegraphics[width=13cm]{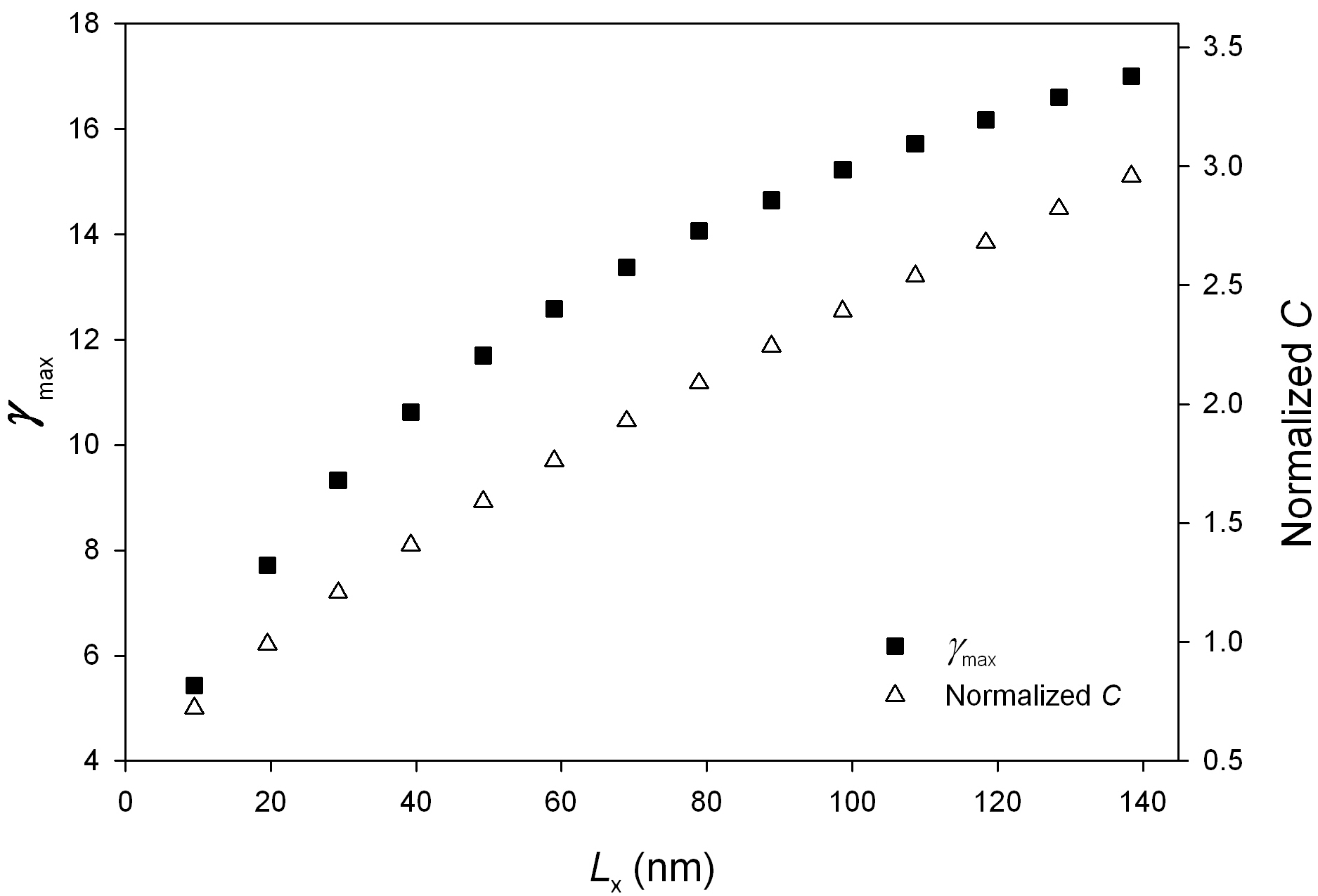}}
\caption{\label{fig:5}
Maximal charge enhancement factor $\gamma_{max}$ and normalized capacitance $C$ of graphene sheets of various lengths $L_{x}$ ($L_{y} \rightarrow \infty$).
} 
\end{figure}

To demonstrate the size effect, we now consider the maximal charge enhancement ratio $\gamma_{max}$ as a function of $L_{x}$. It is shown in Fig. \ref{fig:5} that $\gamma_{max}$ increases with $L_{x}$, the shape of the $\gamma_{max}-vs-L_{x}$ curve is as predicted by the logarithm relationship proposed by Keblinski \textit{et al.} \cite{Keblinski2002}. Since the surface electric potential is a constant, this length dependence directly leads to the rise of the capacitance $C$ of the edge atoms. Also, we plot the normalized $C$ in Fig. \ref{fig:5}. It can be seen that $C$ also increases with $L_{x}$, as predicted by Eq. (\ref{eq:ana23}). However, it is found that the capacitance per unit length decreases with $L_{x}$. This behavior indicates an electric potential loss when the graphene sheet gets shorter, i.e., the longer graphene exhibits a higher surface potential to keep the net charges inside.

\subsection{Multi-layered graphene}
\begin{figure}[ht]
\centerline{\includegraphics[width=13cm]{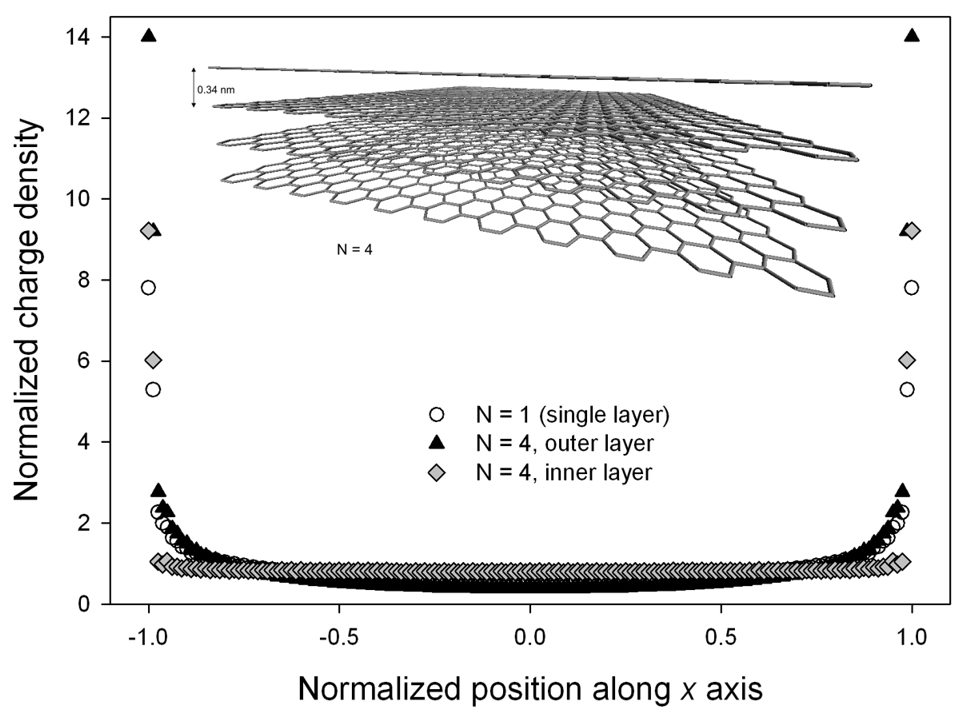}}
\caption{\label{fig:6}
Profile of charge density (normalized to the average) along the $x$ axis in a multi-layered (filled symbols) and a single-layered (empty symbols) graphene sheets ($L_{x}=20$ nm, $L_{y} \rightarrow \infty$).
} 
\end{figure}

Multi-layered graphene sheets have also been studied in this work. They have been reported to have interesting electronic properties.\cite{Hass2008,Orlita2008} We investigated multi-layered graphene in which each layer contains the same quantity of net charges. Due to the electrostatic interaction between the layers, the charge profile in multi-layered graphene cannot be assumed to be a simple addition of that of a single layer. In Fig. \ref{fig:6}, we plot the profile of net charges in a 4-layered graphene to compare with that in  a single-layered one with the same size. It can be seen that the charge enhancement is more significant in multi-layered graphene. The mechanism of this behavior relies on the electrostatic repulsive effects between the carbon layers, which can effectively reduce the capacitance of the atoms at the center of graphene and can however enhance that of the edge atoms.\cite{Zdrojek2008}

\begin{figure}[ht]
\centerline{\includegraphics[width=13cm]{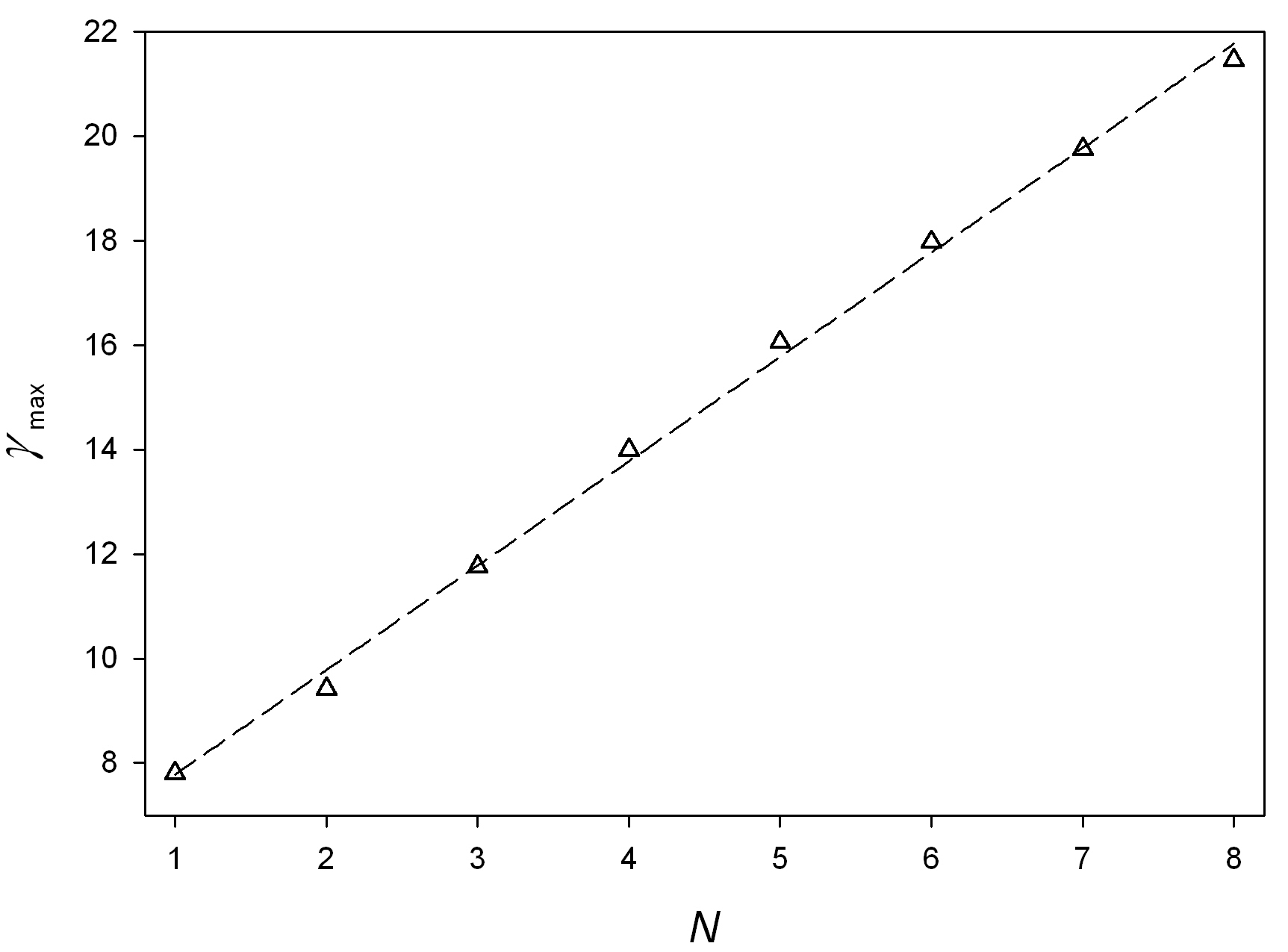}}
\caption{\label{fig:7}
Maximal charge enhancement factor $\gamma_{max}$ vs. the number of layers $N$ in graphene sheets ($L_{x}=20$ nm, $L_{y} \rightarrow \infty$)
} 
\end{figure}

We now estimate the charge enhancement ratio in multi-layered graphene as a function of the number of layer $N$ (Fig. \ref{fig:7}). Our results showed that the charge enhancement effect becomes more significant when the number of layer increases, due to the electrostatic repulsive effect between the layers. We found that $\gamma_{max}$ roughly follows a linear relationship with $N$. Note that the interlayer distance can be slightly changed in case of a very high charge density,\cite{Li2007} while we use a constant distance $0.34$ nm in this work.\cite{Kolmogorov2000}

\newpage

\section{Conclusions}
\label{Sec4}
In conclusion, we have studied the charge distribution in graphene, using both a numerical method and an analytical solution to the classical problem of net charges in a rectangular conducting plate in three physical dimensions. It is found that the analytical model gives a good description to the charge distribution, except for the very edge atoms because of the atomic edge state. We demonstrated how the charge distribution changes with the increasing graphene length, width and number of layer. The length dependence of the charge enhancement effect was found to be related to the rise of graphene capacitance with increasing size. The charge enhancement ratio is found to follow a linear relationship with the number of layers in multi-layered graphene. These results developed a better understanding of the state of net charge in graphene, which can be useful for the potential applications of graphene in charge storages, sensors or field-emission devices. 

\begin{acknowledgments}
We gratefully thank R. Langlet and H.B. Wilson for help with the numerics. A. Mayer, M. Zdrojek, T. Melin, M. Devel and W. Ren are acknowledged for fruitful discussions.
\end{acknowledgments}

\newpage

\section*{Appendix: Variational approximation for the charged rectangular plate}
\label{appdx}

The desired slotion to Laplace's equation Eq. (\ref{eq:ana1}) subject to the symmetric boundary 
conditions Eq. (\ref{eq:ana2}) is even in $x$ and in $y$ (and in $z$), so Fourier cosine integrals
\begin{equation}
\label{eq:ana4}
\psi(x,y,z)=\inthalf d\alpha \inthalf d\beta \, F(\alpha,\beta)\cos(\alpha x)\cos(\beta y)e^{-\gamma|z|}
\end{equation}
are appropriate with $\alpha^2+\beta^2-\gamma^2=0$. The non-analytic modulus $|z|$ accommodates the discontinuity in normal ($z$) derivative 
\begin{equation}
\label{eq:ana6}
\pdd{z}\psi(x,y,0^\pm)=\mp \inthalf d\alpha \inthalf d\beta \, \gamma F(\alpha,\beta)\cos(\alpha x)\cos(\beta y)
\end{equation}
across the surface charge. For $\alpha$ and $\beta$ real, choose $\gamma=+\sqrt{\alpha^2+\beta^2}$.
Using Eq. (\ref{eq:ana7}), Fourier inversion of Eq. (\ref{eq:ana6}) is
\begin{equation}
\label{eq:ana8}
\gamma F(\alpha,\beta)=\left(\frac{2}{\pi}\right)^2k\int\limits_0^{L_x/2} dx \,\frac{\cos(\alpha x)}{\sqrt{1-(2x/L_{x})^2}}\int\limits_0^{L_y/2} dy \,\frac{\cos(\beta y)}{\sqrt{1-(2y/L_{y})^2}} \,\,.
\end{equation}
The required integral is a Bessel function
\begin{equation}
\label{eq:ana9}
\int\limits_0^{L_{x}/2} dx \,\frac{\cos(\alpha x)}{\sqrt{1-(2x/L_{x})^2}}
=\frac{\pi L_{x}}{4}J_0(\alpha L_{x}/2) \,\,.
\end{equation}
Hence the spectral function can be expressed as
\begin{equation}
\label{eq:ana10}
F(\alpha,\beta)=\frac{kL_xL_y}{4}\frac{J_0(\alpha L_x/2)J_0(\beta L_{y}/2)}{\sqrt{\alpha^2+\beta^2}}
\end{equation}
whereupon the forced boundary condition in Eq. (\ref{eq:ana2}) becomes
\begin{equation}
\label{eq:ana11}
\frac{kL_xL_y}{4}\inthalf d\alpha \inthalf d\beta \,\frac{J_0(\alpha L_{x}/2)J_0(\beta L_{y}/2)}{\sqrt{\alpha^2+\beta^2}} \cos(\alpha x)\cos(\beta y)=V_0\qquad 
\begin{array}{c}
\left\{ \begin{array}{l}
|x|<L_{x}/2\\
|y|<L_{y}/2
\end{array}\right.\\
\end{array}\,.
\end{equation}

Application of the Galerkin inner-product operator
\begin{equation}
\int\limits_0^{L_{x}/2} \frac{dx}{\sqrt{1-(2x/L_{x})^2}}\int\limits_0^{L_y/2} \frac{dy}{\sqrt{1-(2y/L_{y})^2}}
\end{equation}
to each side produces
\begin{equation}
\label{eq:ana12}
\frac{kL_x}{2}\inthalf du \inthalf dv \,\frac{J_0^2(u)J_0^2(v)}{\sqrt{u^2+(L_{x}v/L_{y})^2}} =V_0
\end{equation}
where $u=\alpha L_{x}/2$ and $v=\beta L_{y}/2$. With $\mu=L_{x}/L_{y}$, we require the double integral
\begin{equation}
\label{eq:ana13}
I(\mu)=\inthalf du \inthalf dv \,J_0^2(u)J_0^2(v)
\left[u^2+\left(\mu v\right)^2\right]^{-1/2}.
\end{equation}
Note that $I(1/\mu)=\mu I(\mu)$. Using formulae from Watson's treatise\cite{Watson1944}, the square of the Bessel function is expressed as a double integral
\begin{equation}
\label{eq:ana14}
J_0^2(u)=\left(\frac{2}{\pi}\right)^2\int\limits_0^{\pi/2}dx \int\limits_0^{\pi/2}d\xi\,\cos(2u\cos x\cos\xi)
\end{equation}
of trigonometric functions, so that Eq. (\ref{eq:ana13}) temporarily balloons out into the six-fold integral
\begin{equation}
\label{eq:ana15}
I(\mu)=\left(\frac{2}{\pi}\right)^4\int\limits_0^{\pi/2}dx \int\limits_0^{\pi/2}d\xi
\int\limits_0^{\pi/2}dy \int\limits_0^{\pi/2}d\eta
\inthalf du\inthalf dv \,\frac{\cos(2u\cos x\cos\xi)\cos(2v\cos y\cos\eta)}{\bigl[u^2+(\mu v)^2\bigr]^{1/2}}.
\end{equation}
Formula 3.754(2) from Gradshteyn and Ryzhik\cite{Gradshteyn1994} reveals that the $u$--integral is a modified Bessel function
\begin{equation}
\label{eq:ana16}
\inthalf du \,\frac{\cos(2u\cos x\cos\xi)}{\bigl[u^2+(\mu v)^2\bigr]^{1/2}}=K_0(2\mu v\cos x\cos\xi)
\end{equation}
and then their formula 6.671(14) eliminates the Bessel function in the $v$--integral
\begin{equation}
\inthalf dv\,K_0(2\mu v\cos x\cos\xi)=\frac{\pi}{4}\bigl[(\mu\cos x\cos\xi)^2+(\cos y\cos\eta)^2\bigr]^{-1/2}.
\end{equation}
Finally, the original double integral Eq. (\ref{eq:ana13}) of oscillatory functions over infinite ranges is converted to a four-fold integral
\begin{equation}
\label{eq:ana17}
I(\mu)=\frac{4}{\pi^3}\int\limits_0^{\pi/2}dx \int\limits_0^{\pi/2}d\xi
\int\limits_0^{\pi/2}dy \int\limits_0^{\pi/2}d\eta\,
\bigl[(\mu\cos x\cos\xi)^2+(\cos y\cos\eta)^2\bigr]^{-1/2}
\end{equation}
of slowly-varying functions over finite ranges. A specially designed Gaussian quadrature routine
gives numerical values with little fuss. Any one of the four integrals is a complete elliptic integral, but a vectorized four-dimensional quadrature is competitive in execution time
and accuracy with a thee-dimensional quadrature that requires evaluation of the elliptic
function. The desired constant $k$ in Eq. (\ref{eq:ana12}) now takes the form in Eq. (\ref{eq:ana18bis}).

\newpage

\section*{Figures}

\newpage

\newpage

\newpage

\newpage

\newpage

\newpage

\newpage

\newpage

\begin{thebibliography}{56}
\expandafter\ifx\csname natexlab\endcsname\relax\def\natexlab#1{#1}\fi
\expandafter\ifx\csname bibnamefont\endcsname\relax
  \def\bibnamefont#1{#1}\fi
\expandafter\ifx\csname bibfnamefont\endcsname\relax
  \def\bibfnamefont#1{#1}\fi
\expandafter\ifx\csname citenamefont\endcsname\relax
  \def\citenamefont#1{#1}\fi
\expandafter\ifx\csname url\endcsname\relax
  \def\url#1{\texttt{#1}}\fi
\expandafter\ifx\csname urlprefix\endcsname\relax\def\urlprefix{URL }\fi
\providecommand{\bibinfo}[2]{#2}
\providecommand{\eprint}[2][]{\url{#2}}

\bibitem[{\citenamefont{Castro~Neto et~al.}(2009)\citenamefont{Castro~Neto,
  Guinea, Peres, Novoselov, and Geim}}]{CastroNeto2009}
A.H. Castro~Neto, F.~Guinea, N.M.R. Peres, K.S. Novoselov, and A.K. Geim,
  \bibinfo{journal}{Rev. Mod. Phys.} \textbf{\bibinfo{volume}{81}},
  \bibinfo{pages}{109} (\bibinfo{year}{2009}).

\bibitem[{\citenamefont{Beenakker}(2008)}]{Beenakker2008}
C.W.J. Beenakker, \bibinfo{journal}{Rev. Mod. Phys.} \textbf{\bibinfo{volume}{80}},
  \bibinfo{pages}{1337} (\bibinfo{year}{2008}).

\bibitem[{\citenamefont{Hwang et~al.}(2007)\citenamefont{Hwang, Adam, and
  Sarma}}]{Hwang2007d}
E.H. Hwang, S.~Adam, and S. DasSarma,
  \bibinfo{journal}{Phys.\ Rev. Lett.} \textbf{\bibinfo{volume}{98}},
  \bibinfo{pages}{186806} (\bibinfo{year}{2007}).

\bibitem[{\citenamefont{Morozov et~al.}(2008)\citenamefont{Morozov, Novoselov,
  Katsnelson, Schedin, Elias, Jaszczak, and Geim}}]{Morozov2008a}
S.V. Morozov, K.S. Novoselov, M.I. Katsnelson, F.~Schedin, D.C. Elias, J.A.
  Jaszczak, and A.K. Geim,
  \bibinfo{journal}{Phys.\ Rev. Lett.} \textbf{\bibinfo{volume}{100}},
  \bibinfo{pages}{016602} (\bibinfo{year}{2008}).

\bibitem[{\citenamefont{Fogler et~al.}(2008)\citenamefont{Fogler, Guinea, and
  Katsnelson}}]{Fogler2008}
M.M. Fogler, F.~Guinea, and M.I. Katsnelson,
  \bibinfo{journal}{Phys.\ Rev. Lett.} \textbf{\bibinfo{volume}{101}},
  \bibinfo{pages}{226804} (\bibinfo{year}{2008}).

\bibitem[{\citenamefont{Son et~al.}(2006)\citenamefont{Son, Cohen, and
  Louie}}]{Son2006b}
Y.W. Son, M.L. Cohen, and S.G. Louie,
  \bibinfo{journal}{Phys.\ Rev. Lett.} \textbf{\bibinfo{volume}{97}},
  \bibinfo{pages}{216803} (\bibinfo{year}{2006}).

\bibitem[{\citenamefont{Yang et~al.}(2007)\citenamefont{Yang, Park, Son, Cohen,
  and Louie}}]{Yang2007c}
L.~Yang, C.H. Park, Y.W. Son, M.L. Cohen, and S.G. Louie,
  \bibinfo{journal}{Phys.\ Rev. Lett.} \textbf{\bibinfo{volume}{99}},
  \bibinfo{pages}{186801} (\bibinfo{year}{2007}).

\bibitem[{\citenamefont{Han et~al.}(2007)\citenamefont{Han, Ozyilmaz, Zhang,
  and Kim}}]{Han2007a}
M.Y. Han, B.~Ozyilmaz, Y.~Zhang, and P.~Kim,
  \bibinfo{journal}{Phys.\ Rev. Lett.} \textbf{\bibinfo{volume}{98}},
  \bibinfo{pages}{206805} (\bibinfo{year}{2007}).

\bibitem[{\citenamefont{Castro et~al.}(2007)\citenamefont{Castro, Novoselov,
  Morozov, Peres, Dos~Santos, Nilsson, Guinea, Geim, and Neto}}]{Castro2007}
E.V. Castro, K.S. Novoselov, S.V. Morozov, N.M.R. Peres, J.M.B.Lopes dosSantos, Johan Nilsson, F. Guinea, A.K. Geim, and A.H. Castro Neto,
  \bibinfo{journal}{Phys.\ Rev. Lett.} \textbf{\bibinfo{volume}{99}},
  \bibinfo{pages}{216802} (\bibinfo{year}{2007}).

\bibitem[{\citenamefont{Novikov}(2007)}]{Novikov2007a}
D.S. Novikov,
  \bibinfo{journal}{Phys.\ Rev. Lett.} \textbf{\bibinfo{volume}{99}},
  \bibinfo{pages}{056802} (\bibinfo{year}{2007}).

\bibitem[{\citenamefont{Yan et~al.}(2007)\citenamefont{Yan, Zhang, Kim, and
  Pinczuk}}]{Yan2007a}
J.~Yan, Y.~Zhang, P.~Kim, and A.~Pinczuk,
  \bibinfo{journal}{Phys.\ Rev. Lett.} \textbf{\bibinfo{volume}{98}},
  \bibinfo{pages}{166802} (\bibinfo{year}{2007}).

\bibitem[{\citenamefont{Duplock et~al.}(2004)\citenamefont{Duplock, Scheffler,
  and Lindan}}]{Duplock2004}
E.J. Duplock, M.~Scheffler, and P.J.D. Lindan,
  \bibinfo{journal}{Phys.\ Rev. Lett.} \textbf{\bibinfo{volume}{92}},
  \bibinfo{pages}{225502} (\bibinfo{year}{2004}).

\bibitem[{\citenamefont{Biel et~al.}(2009)\citenamefont{Biel, Blase, Triozon,
  and Roche}}]{Biel2009}
B.~Biel, X.~Blase, F.~Triozon, and S.~Roche,
  \bibinfo{journal}{Phys.\ Rev. Lett.} \textbf{\bibinfo{volume}{102}},
  \bibinfo{pages}{096803} (\bibinfo{year}{2009}).

\bibitem[{\citenamefont{Westervelt}(2008)}]{Westervelt2008}
R.M. Westervelt,
  \bibinfo{journal}{Science} \textbf{\bibinfo{volume}{320}},
  \bibinfo{pages}{324} (\bibinfo{year}{2008}).

\bibitem[{\citenamefont{Paillet et~al.}(2005)\citenamefont{Paillet, Poncharal,
  and Zahab}}]{Paillet2005}
M.~Paillet, P.~Poncharal, and A.~Zahab,
  \bibinfo{journal}{Phys.\ Rev. Lett.} \textbf{\bibinfo{volume}{94}},
  \bibinfo{pages}{186801} (\bibinfo{year}{2005}).

\bibitem[{\citenamefont{Zdrojek et~al.}(2005)\citenamefont{Zdrojek, Melin,
  Boyaval, Stivenard, Jouault, Wozniak, Huczko, Gebicki, and
  Adamowicz}}]{Zdrojek2005}
M.~Zdrojek, T.~Melin, C.~Boyaval, D.~Stivenard, B.~Jouault, M.~Wozniak,
  A.~Huczko, W.~Gebicki, and L.~Adamowicz,
  \bibinfo{journal}{Appl. Phys. Lett.} \textbf{\bibinfo{volume}{86}},
  \bibinfo{pages}{213114} (\bibinfo{year}{2005}).

\bibitem[{\citenamefont{Jespersen and Nygard}(2005)}]{Jespersen2005}
T.S. Jespersen and J.~Nygard,
  \bibinfo{journal}{Nano Lett.} \textbf{\bibinfo{volume}{5}},
  \bibinfo{pages}{1838} (\bibinfo{year}{2005}).

\bibitem[{\citenamefont{Charlier et~al.}(2007)\citenamefont{Charlier, Blase,
  and Roche}}]{Charlier2007}
J.C. Charlier, X.~Blase, and S.~Roche,
  \bibinfo{journal}{Rev. Mod. Phys.} \textbf{\bibinfo{volume}{79}},
  \bibinfo{pages}{677} (\bibinfo{year}{2007}).

\bibitem[{\citenamefont{Morozov et~al.}(2006)\citenamefont{Morozov, Novoselov,
  Katsnelson, Schedin, Ponomarenko, Jiang, and Geim}}]{Morozov2006}
S.V. Morozov, K.S. Novoselov, M.I. Katsnelson, F.~Schedin, L.A. Ponomarenko,
  D.~Jiang, and A.K. Geim,
  \bibinfo{journal}{Phys.\ Rev. Lett.} \textbf{\bibinfo{volume}{97}},
  \bibinfo{pages}{016801} (\bibinfo{year}{2006}).

\bibitem[{\citenamefont{Yan and Ting}(2008)}]{Yan2008}
X.Z. Yan and C.S. Ting,
  \bibinfo{journal}{Phys.\ Rev. Lett.} \textbf{\bibinfo{volume}{101}},
  \bibinfo{pages}{126801} (\bibinfo{year}{2008}).

\bibitem[{\citenamefont{Rossi and Das~Sarma}(2008)}]{Rossi2008}
E.~Rossi and S.~Das~Sarma,
  \bibinfo{journal}{Phys.\ Rev. Lett.} \textbf{\bibinfo{volume}{101}},
  \bibinfo{pages}{166803} (\bibinfo{year}{2008}).

\bibitem[{\citenamefont{Cui et~al.}(2002)\citenamefont{Cui, Sordan, Burghard,
  and Kern}}]{Cui2002}
J.B. Cui, R.~Sordan, M.~Burghard, and K.~Kern,
  \bibinfo{journal}{Appl. Phys. Lett.} \textbf{\bibinfo{volume}{81}},
  \bibinfo{pages}{3260} (\bibinfo{year}{2002}).

\bibitem[{\citenamefont{Robinson et~al.}(2008)\citenamefont{Robinson, Perkins,
  Snow, Wei, and Sheehan}}]{Robinson2008}
J.T. Robinson, F.K. Perkins, E.S. Snow, Z.~Wei, and P.E. Sheehan,
  \bibinfo{journal}{Nano Lett.} \textbf{\bibinfo{volume}{8}},
  \bibinfo{pages}{3137} (\bibinfo{year}{2008}).

\bibitem[{\citenamefont{Bonard et~al.}(2002)\citenamefont{Bonard, Dean, Coll,
  and Klinke}}]{Bonard2002}
J.M. Bonard, K.A. Dean, B.F. Coll, and C.~Klinke,
  \bibinfo{journal}{Phys.\ Rev. Lett.} \textbf{\bibinfo{volume}{89}},
  \bibinfo{pages}{197602} (\bibinfo{year}{2002}).

\bibitem[{\citenamefont{Purcell
  et~al.}(2002{\natexlab{b}})\citenamefont{Purcell, Vincent, Journet, and
  Binh}}]{Purcell2002}
S.T. Purcell, P.~Vincent, C.~Journet, and V.~ThienBinh,
  \bibinfo{journal}{Phys.\ Rev. Lett.} \textbf{\bibinfo{volume}{88}},
  \bibinfo{pages}{105502} (\bibinfo{year}{2002}{\natexlab{b}}).

\bibitem[{\citenamefont{Maxwell}(1873)}]{Maxwell1873}
\bibinfo{author}{\bibfnamefont{J.~C.} \bibnamefont{Maxwell}},
  \emph{\bibinfo{title}{A Treatise on Electricity and Magnetism}}
  (\bibinfo{publisher}{Clarendon Press (1891)}, \bibinfo{year}{1873}).

\bibitem[{\citenamefont{Jackson}(1975)}]{Jacksonbook1975}
\bibinfo{author}{\bibfnamefont{J.~D.} \bibnamefont{Jackson}},
  \emph{\bibinfo{title}{Classical Electrodynamics}} (\bibinfo{publisher}{Wiley,
  New York}, \bibinfo{year}{1975}), \bibinfo{note}{p. 57-94}.

\bibitem[{\citenamefont{Morse and Feshbach}(1953)}]{Morse1953}
\bibinfo{author}{\bibfnamefont{P.~M.} \bibnamefont{Morse}} \bibnamefont{and}
  \bibinfo{author}{\bibfnamefont{H.}~\bibnamefont{Feshbach}},
  \emph{\bibinfo{title}{Methods of Theoretical Physics: II}}
  (\bibinfo{publisher}{McGraw-Hill (New York)}, \bibinfo{year}{1953}),
  \bibinfo{note}{p. 1196-1197}.

\bibitem[{\citenamefont{Bottcher}(1952)}]{Bottcher1952}
\bibinfo{author}{\bibfnamefont{C.J.F.}~\bibnamefont{Bottcher}},
  \emph{\bibinfo{title}{Theory of electric polarization}}
  (\bibinfo{publisher}{Elsevier (Amsterdam)}, \bibinfo{year}{1952}),
  \bibinfo{note}{p. 48-60}.

\bibitem[{\citenamefont{Mayer}(2007)}]{mayer-07-01}
\bibinfo{author}{\bibfnamefont{A.}~\bibnamefont{Mayer}},
  \bibinfo{journal}{Phys.\ Rev.~B} \textbf{\bibinfo{volume}{75}},
  \bibinfo{pages}{045407} (\bibinfo{year}{2007}).

\bibitem[{\citenamefont{Wang}(2009)}]{ZW2009}
\bibinfo{author}{\bibfnamefont{Z.}~\bibnamefont{Wang}},
  \bibinfo{journal}{Phys.\ Rev.~B} \textbf{\bibinfo{volume}{79}},
  \bibinfo{pages}{155407} (\bibinfo{year}{2009}).

\bibitem[{\citenamefont{Applequist et~al.}(1972)\citenamefont{Applequist, Carl,
  and Fung}}]{Applequist-72}
J. Applequist, J.R. Carl, and K.K. Fung,
  \bibinfo{journal}{J. Am. Chem. Soc.} \textbf{\bibinfo{volume}{94}},
  \bibinfo{pages}{2952} (\bibinfo{year}{1972}).

\bibitem[{\citenamefont{Olson and Sundberg}(1978)}]{olson-78}
M.L. Olson and K.R. Sundberg,
  \bibinfo{journal}{J. Chem. Phys.} \textbf{\bibinfo{volume}{69}},
  \bibinfo{pages}{5400} (\bibinfo{year}{1978}).

\bibitem[{\citenamefont{Jensen, L.}(2002)}]{Jensen2002}
L. Jensen, P.O. Astrand, A. Osted, J. Kongsted and K.V. Mikkelsen,
  \bibinfo{journal}{J. Chem. Phys.} \textbf{\bibinfo{volume}{116}},
  \bibinfo{pages}{4001} (\bibinfo{year}{2002}).

\bibitem[{\citenamefont{Mayer}(2005{\natexlab{a}})}]{mayer-05-01}
\bibinfo{author}{\bibfnamefont{A.}~\bibnamefont{Mayer}},
  \bibinfo{journal}{Phys.\ Rev.~B} \textbf{\bibinfo{volume}{71}},
  \bibinfo{pages}{235333} (\bibinfo{year}{2005}{\natexlab{a}}).

\bibitem[{\citenamefont{Mayer}(2005{\natexlab{b}})}]{Mayer2005c}
\bibinfo{author}{\bibfnamefont{A.}~\bibnamefont{Mayer}},
  \bibinfo{journal}{Appl. Phys. Lett.} \textbf{\bibinfo{volume}{86}},
  \bibinfo{pages}{153110} (\bibinfo{year}{2005}{\natexlab{b}}).

\bibitem[{\citenamefont{Wang et~al.}(2008)\citenamefont{Wang, Zdrojek,
  M\'{e}lin, and Devel}}]{zhaowang-08-01}
\bibinfo{author}{\bibfnamefont{Z.}~\bibnamefont{Wang}},
  \bibinfo{author}{\bibfnamefont{M.}~\bibnamefont{Zdrojek}},
  \bibinfo{author}{\bibfnamefont{T.}~\bibnamefont{M\'{e}lin}},
  \bibnamefont{and} \bibinfo{author}{\bibfnamefont{M.}~\bibnamefont{Devel}},
  \bibinfo{journal}{Phys.\ Rev.~B} \textbf{\bibinfo{volume}{78}},
  \bibinfo{pages}{085425} (\bibinfo{year}{2008}).

\bibitem[{\citenamefont{Stuart et~al.}(2000)\citenamefont{Stuart, Tutein, and
  Harrison}}]{Stuart2000a}
S.J. Stuart, A.B. Tutein, and J.A. Harrison,
  \bibinfo{journal}{J. Chem. Phys.} \textbf{\bibinfo{volume}{112}},
  \bibinfo{pages}{6472} (\bibinfo{year}{2000}).

\bibitem[{\citenamefont{Ni et~al.}(2002)\citenamefont{Ni, Sinnott, Mikulski,
  and Harrison}}]{Ni2002}
B.~Ni, S.B. Sinnott, P.T. Mikulski, and J.A. Harrison,
  \bibinfo{journal}{Phys.\ Rev. Lett.} \textbf{\bibinfo{volume}{88}},
  \bibinfo{pages}{205505} (\bibinfo{year}{2002}).

\bibitem[{\citenamefont{Shenoy et~al.}(2008)\citenamefont{Shenoy, Reddy,
  Ramasubramaniam, and Zhang}}]{Shenoy2008}
V.B. Shenoy, C.D. Reddy, A.~Ramasubramaniam, and Y.W. Zhang,
  \bibinfo{journal}{Phys.\ Rev. Lett.} \textbf{\bibinfo{volume}{101}},
  \bibinfo{pages}{245501} (\bibinfo{year}{2008}).

\bibitem[{\citenamefont{Wang and Philippe}()}]{ZW200903}
\bibinfo{author}{\bibfnamefont{Z.}~\bibnamefont{Wang}} \bibnamefont{and}
  \bibinfo{author}{\bibfnamefont{L.}~\bibnamefont{Philippe}},
  \bibinfo{journal}{Phys.\ Rev. Lett.} \textbf{\bibinfo{volume}{102}},
  \bibinfo{pages}{215501} (\bibinfo{year}{2009}).

\bibitem[{\citenamefont{Wang and Devel}(2007)}]{zhaowang-07-03}
\bibinfo{author}{\bibfnamefont{Z.}~\bibnamefont{Wang}} \bibnamefont{and}
  \bibinfo{author}{\bibfnamefont{M.}~\bibnamefont{Devel}},
  \bibinfo{journal}{Phys.\ Rev.~B} \textbf{\bibinfo{volume}{76}},
  \bibinfo{pages}{195434} (\bibinfo{year}{2007}).

\bibitem[{\citenamefont{Wang et~al.}(2007)\citenamefont{Wang, Devel, Langlet,
  and Dulmet}}]{zhaowang-07-02}
\bibinfo{author}{\bibfnamefont{Z.}~\bibnamefont{Wang}},
  \bibinfo{author}{\bibfnamefont{M.}~\bibnamefont{Devel}},
  \bibinfo{author}{\bibfnamefont{R.}~\bibnamefont{Langlet}}, \bibnamefont{and}
  \bibinfo{author}{\bibfnamefont{B.}~\bibnamefont{Dulmet}},
  \bibinfo{journal}{Phys.\ Rev.~B} \textbf{\bibinfo{volume}{75}},
  \bibinfo{pages}{205414} (\bibinfo{year}{2007}).

\bibitem[{\citenamefont{Gotoi}(1992)}]{Gotoi1992}
\bibinfo{author}{\bibfnamefont{E.}~\bibnamefont{Gotoi}}, \bibinfo{journal}{J.
  Comp. Phys.} \textbf{\bibinfo{volume}{100}}, \bibinfo{pages}{105}
  (\bibinfo{year}{1992}).

\bibitem[{\citenamefont{Keblinski et~al.}(2002)\citenamefont{Keblinski, Nayak,
  Zapol, and Ajayan}}]{Keblinski2002}
P.~Keblinski, S.K. Nayak, P.~Zapol, and P.M. Ajayan,
  \bibinfo{journal}{Phys.\ Rev. Lett.} \textbf{\bibinfo{volume}{89}},
  \bibinfo{pages}{255503} (\bibinfo{year}{2002}).

\bibitem[{\citenamefont{Bonard et~al.}(1998)\citenamefont{Bonard, Stockli,
  Maier, de~Heer, Chatelain, Salvetat, and Forro}}]{Bonard1998}
J.M. Bonard, T.~Stockli, F.~Maier, W.A. de~Heer, A.~Chatelain, J.P. Salvetat,
  and L.~Forro,
  \bibinfo{journal}{Phys.\ Rev. Lett.} \textbf{\bibinfo{volume}{81}},
  \bibinfo{pages}{1441} (\bibinfo{year}{1998}).

\bibitem[{\citenamefont{Liu et~al.}(2009)\citenamefont{Liu, Suenaga, Harris,
  and Iijima}}]{Liu2009}
Z.~Liu, K.~Suenaga, P.J.F. Harris, and S.~Iijima,
  \bibinfo{journal}{Phys.\ Rev. Lett.} \textbf{\bibinfo{volume}{102}},
  \bibinfo{pages}{015501} (\bibinfo{year}{2009}).

\bibitem[{\citenamefont{Yang et~al.}(2008)\citenamefont{Yang, Cohen, and
  Louie}}]{Yang2008d}
L.~Yang, M.L. Cohen, and S.G. Louie,
  \bibinfo{journal}{Phys.\ Rev. Lett.} \textbf{\bibinfo{volume}{101}},
  \bibinfo{pages}{186401} (\bibinfo{year}{2008}).

\bibitem[{\citenamefont{Abanin et~al.}(2006)\citenamefont{Abanin, Lee, and
  Levitov}}]{Abanin2006}
D.A. Abanin, P.A. Lee, and L.S. Levitov,
  \bibinfo{journal}{Phys.\ Rev. Lett.} \textbf{\bibinfo{volume}{96}},
  \bibinfo{pages}{176803} (\bibinfo{year}{2006}).

\bibitem[{\citenamefont{Hass et~al.}(2008)\citenamefont{Hass, Varchon,
  Millan-Otoya, Sprinkle, Sharma, de~Heer, Berger, First, Magaud, and
  Conrad}}]{Hass2008}
J.~Hass, F.~Varchon, J.E. Millan-Otoya, M.~Sprinkle, N.~Sharma, W.A. de~Heer,
  C.~Berger, P.N. First, L.~Magaud, and E.H. Conrad,
  \bibinfo{journal}{Phys.\ Rev. Lett.} \textbf{\bibinfo{volume}{100}},
  \bibinfo{pages}{125504} (\bibinfo{year}{2008}).

\bibitem[{\citenamefont{Orlita et~al.}(2008)\citenamefont{Orlita, Faugeras,
  Plochocka, Neugebauer, Martinez, Maude, Barra, Sprinkle, Berger, de~Heer
  et~al.}}]{Orlita2008}
M.~Orlita, C.~Faugeras, P.~Plochocka, P.~Neugebauer, G.~Martinez, D.K. Maude,
  A.L. Barra, M.~Sprinkle, C.~Berger, W.A. de~Heer, and M.~Potemski, \bibinfo{journal}{Phys.\ Rev. Lett.}
  \textbf{\bibinfo{volume}{101}}, \bibinfo{pages}{267601}
  (\bibinfo{year}{2008}).

\bibitem[{\citenamefont{Zdrojek et~al.}(2008)\citenamefont{Zdrojek, Heim,
  Brunel, Mayer, and M\'{e}lin}}]{Zdrojek2008}
M.~Zdrojek, T.~Heim, D.~Brunel, A.~Mayer, and T.~M\'{e}lin,
  \bibinfo{journal}{Phys.\ Rev.~B} \textbf{\bibinfo{volume}{77}},
  \bibinfo{pages}{033404} (\bibinfo{year}{2008}).

\bibitem[{\citenamefont{Li and Chou}(2007)}]{Li2007}
C.~Li and T.W. Chou,
  \bibinfo{journal}{Carbon} \textbf{\bibinfo{volume}{45}}, \bibinfo{pages}{922}
  (\bibinfo{year}{2007}).

\bibitem[{\citenamefont{Kolmogorov and Crespi}(2000)}]{Kolmogorov2000}
A.N. Kolmogorov and V.H. Crespi,
  \bibinfo{journal}{Phys.\ Rev. Lett.} \textbf{\bibinfo{volume}{85}},
  \bibinfo{pages}{4727} (\bibinfo{year}{2000}).

\bibitem[{\citenamefont{Watson}(1944)}]{Watson1944}
G.N. Watson,
  \emph{\bibinfo{title}{A Treatise on the Theory of Bessel Functions}}
  (\bibinfo{publisher}{Cambridge University Press, 2nd Edition},
  \bibinfo{year}{1944}),  \bibinfo{note}{p. 48, p. 151}.

\bibitem[{\citenamefont{Gradshteyn and Ryzhik}(1994)}]{Gradshteyn1994}
I.S. Gradshteyn and I.M. Ryzhik,
  \emph{\bibinfo{title}{Table of Integrals, Series, and Products}}
  (\bibinfo{publisher}{Academic Press (Boston), 5th Edition},
  \bibinfo{year}{1994}).

\bibitem[{\citenamefont{HarringtonMM}(2009)}]{HarringtonMM}
R.F. Harrington,
  \emph{\bibinfo{title}{Field Computation by Moment Methods}}
  (\bibinfo{publisher}{Macmillan (New York)},
  \bibinfo{year}{1968}),  \bibinfo{note}{p. 24-28}.


\end{thebibliography}
\end{document}